\documentclass[aps,prb,twocolumn,amsmath,amssymb,showpacs,superscriptaddress]{revtex4-1}

\usepackage{graphicx}
\usepackage{color}

\makeatletter

\@ifundefined{textcolor}{}
{
 \definecolor{BLACK}{gray}{0}
 \definecolor{WHITE}{gray}{1}
 \definecolor{RED}{rgb}{1,0,0}
 \definecolor{GREEN}{rgb}{0,1,0}
 \definecolor{BLUE}{rgb}{0,0,1}
 \definecolor{CYAN}{cmyk}{1,0,0,0}
 \definecolor{MAGENTA}{cmyk}{0,1,0,0}
 \definecolor{YELLOW}{cmyk}{0,0,1,0}
}

\begin{document}

%
\title{The effects of Moir\'e lattice on the electronic properties of graphene.}
%
%
%
\author{Lunan Huang}
\author{Yun Wu}
\author{M. T. Hershberger}
\author{Daixiang Mou}
\author{Ben Schrunk}
\author{Michael C. Tringides}
\author{Myron Hupalo}
\author{Adam Kaminski}
\affiliation{Division of Materials Science and Engineering, Ames Laboratory, U.S. DOE, Ames, Iowa 50011, USA}
\affiliation{Department of Physics and Astronomy, Iowa State University, Ames, Iowa 50011, USA}


\begin{abstract}
We study structural and electronic properties of graphene grown on SiC substrate using scanning tunneling microscope (STM), spot-profile-analysis low energy electron diffraction (SPA-LEED) and angle resolved photoemission spectroscopy (ARPES). 
We find several new replicas of Dirac cones in the Brillouin zone (BZ). 
Their locations can be understood in terms of combination of basis vectors linked to SiC 6 $\times$ 6 and graphene $6\sqrt{3} \times 6\sqrt{3}$ reconstruction. 
Therefore these new features originate from the Moir\'e caused by the  lattice mismatch between SiC and graphene. 
More specifically, Dirac cones replicas are caused by underlying weak modulation of the ionic potential by the substrate that is then experienced by the electrons in the graphene. 
We also demonstrate that this effect is equally strong in single and tri-layer graphene, therefore the additional Dirac cones are intrinsic features rather than result of photoelectron diffraction. 
These new features in the electronic structure are very important for the interpretation of recent  transport measurements and can assist  in tuning the properties of graphene for practical applications. 
\end{abstract}


\maketitle

\section{Introduction}
In the last decade, graphene has become a topic of intense research because of its unique structural and electronic properties such as presence of Dirac dispersion, which leads to high thermal conductivity \cite{doi:10.1021/nl0731872}, ballistic transport \cite{Du2008}, and ultrahigh electron mobility \cite{Bolotin2008351}.
Each carbon atom has a $\pi$-bond perpendicular to the graphene plane. These bonds are jointly  hybridized  to form $\pi$-bands and $\pi$* bands\cite{Cooper2012}. 
Graphene can be readily grown on large area insulating, semiconducting and metallic substrates. 
Lattice mismatch at the substrate interface leads to the formation of Moir\'e patterns (i.e. a superlattice) with larger periodicity than the  lattice constants of each of the two separate  lattices. 
This is a standard way to obtain weak periodic potentials superimposed on the graphene potential with characteristic periodicity of several nanometers. 
The strain caused by lattice mismatch is one of important ways to tune the graphene electronic structure and properties \cite{Park2008a, Park2008, Ponomarenko2013, Decker2011}. 
Recently, tuning  the periodicity of the ionic potential experienced by electrons in graphene grown on boron nitride substrate  and measuring its fractional quantum Hall resistance enabled the engineering of the energy spectrum of Hofstadter butterfly \cite{Hofstadter1976,Dean2013,Hunt2013,Yu2014,Wang2015}.

Silicone carbide (SiC) is one of the most common substrates to grow graphene because of its hexagonal crystal structure with lattice constant a$_{SiC}$ = b$_{SiC}$ = 3.073 \AA, c = 10.053 \AA. 
The 6H-SiC consists of Si-C bilayers that are stacked  in the pattern of ABCACB\cite{0953-8984-20-32-323202}.  Because of  different lattice constants of SiC and graphene, when graphene is grown, an intermediate layer will be constructed at the SiC surface between substrate and graphene. This intermediate layer may exhibit rotational disorder, forming Moir\'e patterns\cite{VanBommel1975463, PhysRevB.58.16396}.  

Epitaxial graphene grown by thermal annealing of SiC has been studied extensively with several complementary techniques to reveal its structural and electronic properties. 
These studies helped to better understand many aspects of graphene layer on SiC (ionic center position, thickness uniformity, stacking, relative layer orientation and variation of the band structure with number of graphene layers) \cite{VanBommel1975463,starke1997large,PhysRevB.58.16396,PhysRevB.77.155426,0953-8984-20-32-323202,Xu2014,PhysRevLett.98.206802}. 
However, a number of questions still remain about the nature of the graphene-substrate interface and how it affects the Dirac fermions. 
The layer at the interface, referred to as the ``buffer" or ``zeroth" layer graphene, has no $\pi$-bands\cite{PhysRevLett.99.126805}. 
This layer increases the carrier concentration and shifts the Femi level, without modifying the shape of the Dirac cones \cite{PhysRevLett.99.126805}.
Structurally, the buffer layer was represented in terms of the two coincidence lattices, which form two distinct diffraction patterns: (1) 6 $\times$ 6 (oriented along the SiC unit cell) and (2) $6\sqrt{3} \times 6\sqrt{3}$ rotated 30$^\cdot$ from the 6 $\times$ 6 unit cell. 
A new type of buffer layer was grown with linear $\pi$-bands separated by a measurable gap \cite{Zhou2007}.
This study motivated new experiments to correlate structural and electronic information necessary for understanding and controlling the properties of graphene. 
The surface reconstruction has been studied by several techniques in the past, including LEED \cite{VanBommel1975463, PhysRevB.58.16396, starke1997large}, auger electron spectroscopy \cite{VanBommel1975463, starke1997large}, scanning tunneling microscope \cite{starke1997large, Xu2014} and ARPES \cite{PhysRevLett.98.206802, Bostwick2006, Ohta951, 1367-2630-9-10-385, Mucha-Kruczyfmmodecutenlseniski2008, 1367-2630-10-2-023034, PhysRevLett.101.086402, PhysRevB.83.121408, Walter2011, Ohta2012, Nevius2015}. Coincidence lattices have been used routinely to probe buried interfaces and deduce information about geometric or electronic corrugations\cite{Emtsev2008}. 

\begin{figure*}
	\centering
	\includegraphics[width=7in]{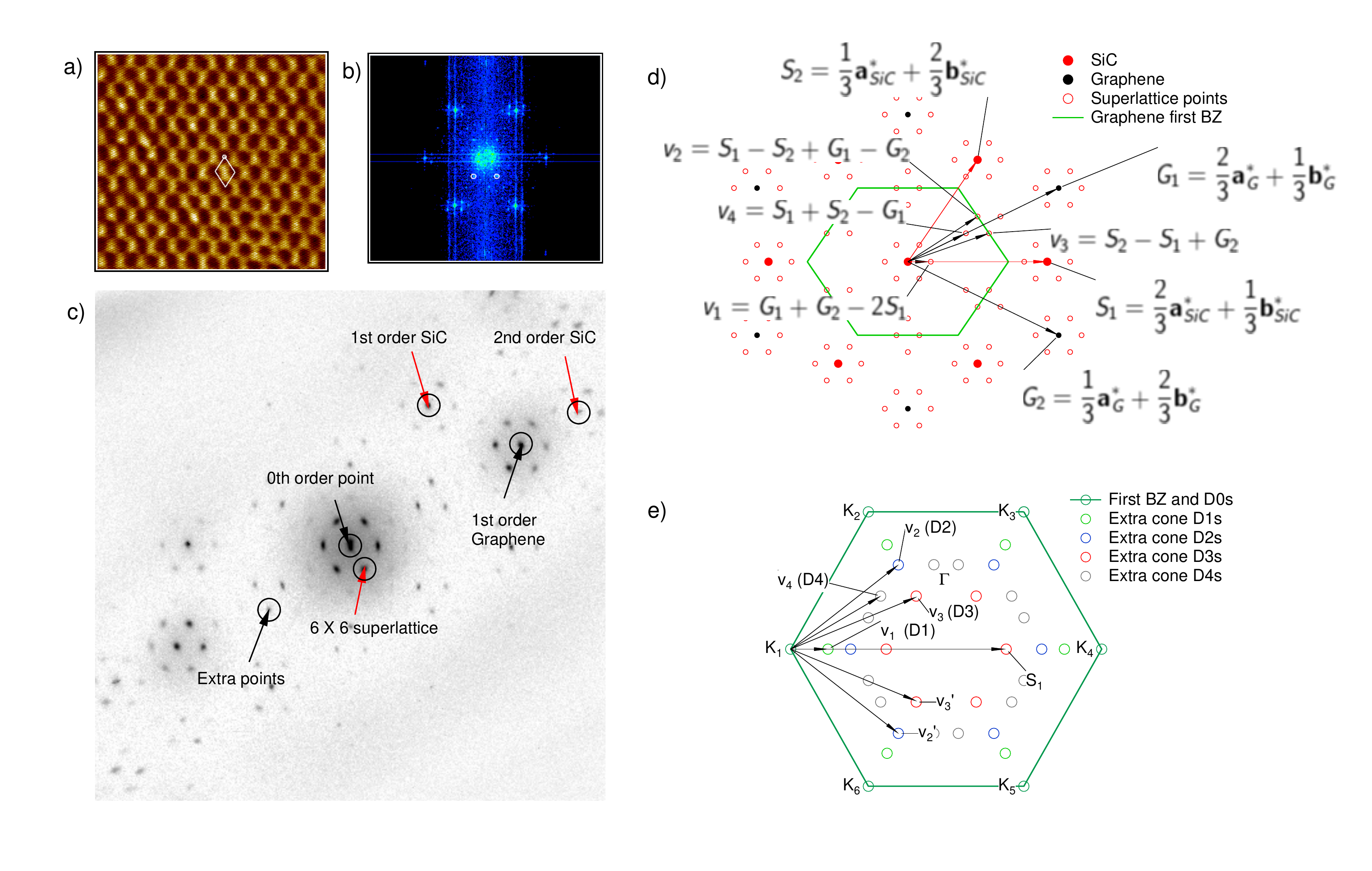}
	\caption{
	(a) STM image of graphene grown on 6H-SiC substrate. The size of the scan was 20 nm x 18 nm. 
	(b) Fourier transform of (a), white circles mark locations of Moir\'e peaks. 
	(c) SPA-LEED pattern. 
	(d) sketch of diffraction patterns in (c). The relevant vectors are marked by arrows. 
	(e) Sketch of expected locations of the replica Dirac cones based on (d) $a^*_{SiC}$, $b^*_{SiC}$ are the reciprocal primitive vectors of SiC and $a^*_G$, $b^*_G$ are the reciprocal primitive vectors of graphene.}
	\label{fig:fig1}
\end{figure*}

The structure of the buffer layer is still an open question. 
The first detailed work explored this using LEED and STM and proposed a  commensurate model  based on the Si-dangling bonds \cite{Emtsev2008}. 
The electronic structure of this model was predicted to be  insulating. 
Recent work \cite{Ohta2012} has shown that the buffer layer can be semiconducting if the right annealing conditions are followed with a small gap opening up. 
This was shown to be related to a hidden incommensurate structure present in the buffer layer  and that the structure of the buffer layer is strongly coupled to the SiC bilayer\cite{Conrad2017}.  
Additional ARPES and STM work has mapped out the initial stage of the buffer layer and the corresponding changes to the band structure. 
The work  has shown that the small gap is seen to evolve from the initial electronic features of the  buffer layer  to the semiconducting phase with the 6x6 SiC periodicity visible in constant energy ARPES maps \cite{Nair2017}.
Our work shows the same three Dirac replica cones  for both monolayer and trilayer. 
The cones are defined by the same 6x6 sublattice reciprocal lattice vectors  so it is consistent and complement the previous measurements \cite{Rotenberg2015}. 
It is very puzzling why even after the graphene layer is complete the key remaining wavevectors are the 6x6 and not the 13x13 despite the graphene layer becoming  uniform. It was also shown \cite{Riedl2009} that hydrogen intercalation of decouples the graphene layer from the substrate.  

Here we focus on the 6x6 reconstruction that marks the onset of formation of the buffer layer and it remains present even after additional graphene layers are grown.  This reconstruction leads to appearance of new features in the electronic structure, namely additional replicas of the Dirac cones. 
The photoelectron intensity of these objects does not decrease with increasing number of graphene layers and each replica has very different pattern of intensities rapidly changing with momentum and binding energy. This  demonstrates their intrinsic origin, rather than trivial photoelectron diffraction. 
In fact, the pattern of the band dispersion within these replica cones and the corresponding  wavevectors they appear at, prove that they arise due to weak modulation of the electronic potential of graphene. The modulation  is  caused by the interplay of lattice periodicities, i.~e. formation of Moir\'e pattern. 
This modulation of the graphene potential is very relevant to recent transport results and provides a pathway for understanding its properties. 

\section{Experimental Details}
We use commercial 6H-SiC substrate to grow graphene. 
We first anneal the 2$\times$12 mm substrate at 600 $^{\circ}$C for 3 hours to clean its surface and then increase the temperature of the SiC to 1200 $^{\circ}$C for 10 minutes. 
This procedure results in a growth of a single layer of graphene, as confirmed by STM, LEED and ARPES. 
Tri-layer graphene is grown by heating up the same sample for another 10 minutes at 1200 $^{\circ}$C. ARPES measurements were performed at Ames Laboratory using a high precision ARPES spectrometer that consists of a Scienta SES2002 electron analyzer and a GammaData Helium UV lamp equipped with custom designed refocusing optics. 
All data were acquired using the HeI line with a photon energy of 21.2 eV. 
The angular resolution was 0.13$^\circ$ and $\sim$ 0.5$^\circ$ along and perpendicular to the direction of the analyzer slits, respectively. The energy corresponding to the chemical potential was determined from the Fermi edge of a polycrystalline Au reference in electrical contact with the sample. The energy resolution was set at $\sim$20meV - confirmed by measuring the energy width between 90\% and 10\% of the Fermi edge from the same Au reference. The data were measured using several samples yielding consistent results.

\begin{table}
	\begin{tabular}{| l | l | l | l |}
		\hline
		& D1 & D2 & D3 \\ \hline
		Calculated position (in units of $\Gamma$K distance) & 0.230 & 0.384 & 0.615 \\ \hline
		Measured position & 0.240 & 0.390 & 0.619\\ \hline
		Error (\%) & 4.3 & 1.6 & 0.7 \\ \hline
	\end{tabular}
	\caption{Comparison of calculated and measured wavevectors of the Dirac cones.}
	\label{table1}
\end{table}

\begin{figure}
	\centering
	\includegraphics[width=\columnwidth]{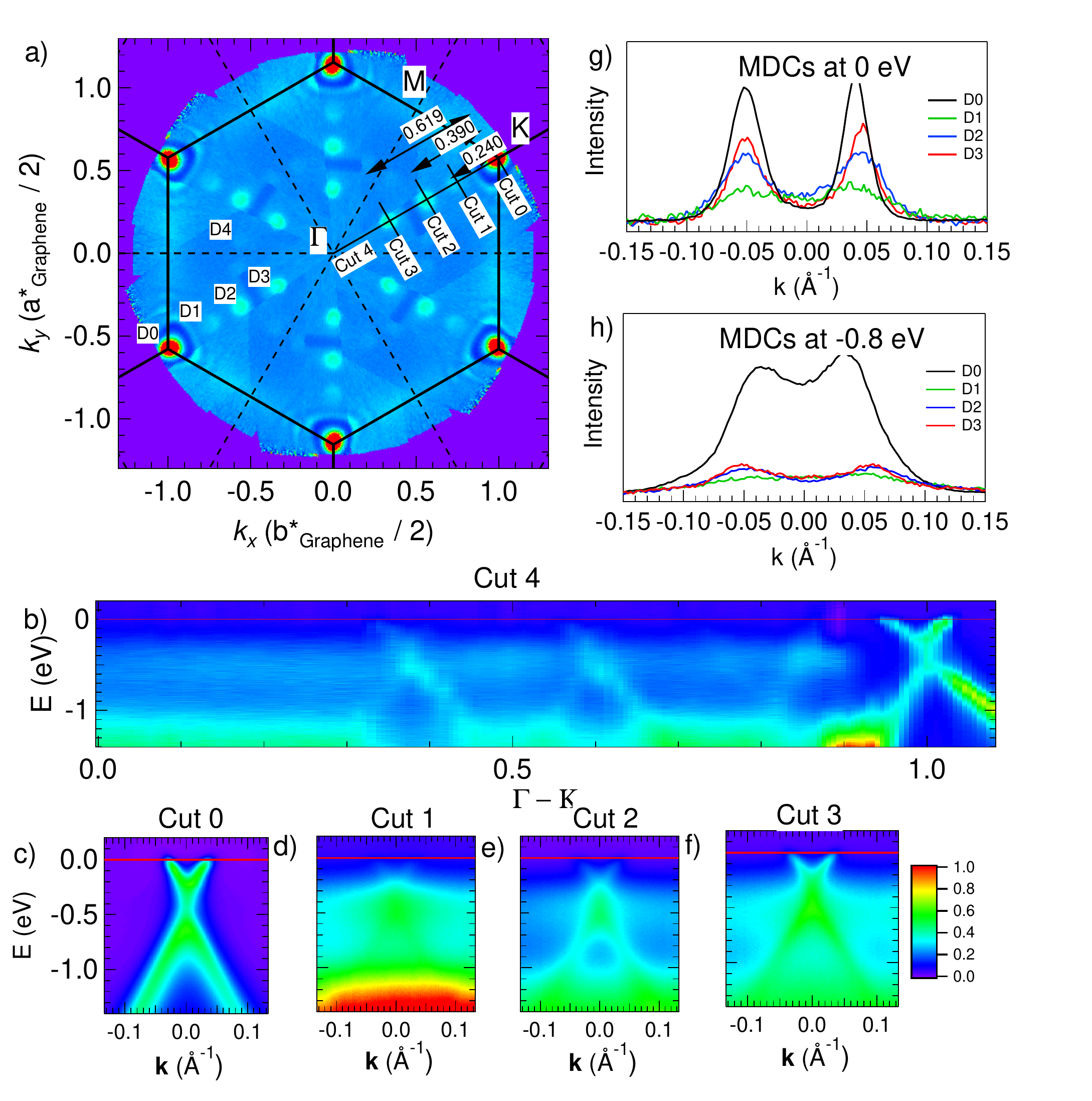}
	\caption{
	(a) Fermi surface of monolayer graphene grown on 6H-SiC substrate. 
	(b) Energy dispersion along $\Gamma$ - K direction. 
	(c - f) Energy dispersion of Dirac cones D0, D1, D2 and D3 (along directions marked as Cut 0 - 3 in (a).
	(g) MDCs at Fermi energy from (c - f). 
	(h) MDCs at 0.8 eV below E$_F$ from (c - f).}
	\label{fig:fig2}
\end{figure}

\section{Results and Discussion}
Figure \ref{fig:fig1}a shows the topography of graphene layer measured by STM. Clear Moir\'e pattern is visible as periodic ``checkerboard"-like pattern. 
The intensity variations are due to the combination of the periodic changes in layer height and electronic densities.
A small rhombus is used to outline the 6 $\times$ 6 ``quasi cell". 
Figure \ref{fig:fig1}b shows the Fourier transform of \ref{fig:fig1}a. 
The bright points in white circles are due to the $6\sqrt{3} \times 6\sqrt{3}$ lattice modulation. 
This data is consistent with the result of SPA-LEED shown in Figure \ref{fig:fig1}c. 
The zero order spot is located at the center of the image and surrounded by six ``$6 \times 6$" spots. 
The first order diffraction peaks from SiC are surrounding the center peak with smaller radius. 
The graphene first order peaks are further away and rotated by 30$^\circ$ from the SiC pattern. 
Each of the diffraction peaks is surrounded by six ``$6 \times 6$" spots as expected. 
The ratios of the positions of graphene and SiC first diffraction spots is 1.26, are consistent with ratios of their respective reciprocal lattice constants (3.08 : 2.46). 
In between zeroth and first order diffraction peaks of graphene there are four additional, weaker peaks due to the $6\sqrt{3} \times 6\sqrt{3}$ lattice modulation. 
A schematic drawing of all observed diffraction peaks is shown in Figure \ref{fig:fig1}d based on data in panels (b) and (c). 
Vectors S$_1$ and S$_2$ are pointing to SiC reciprocal points, while vectors G$_1$ and G$_2$ are of graphene layer. 
All other points arise due to combinations of the S and G vectors. 
For example, vector $v_1$ is obtained as G$_1$ + G$_2$ - 2S$_1$ and points to one of the six satellite peaks around the center. 
In defining the four vectors $v_1$, $v_2$, $v_3$, $v_4$ it is important to define the lattice constants of the two unit cells of the 6 $\times$ 6 and $6\sqrt{3} \times \sqrt{3}$ coincidence lattices. 
If we use the graphene BZ (BZ = $\frac{2\pi}{a_G}$) as 100$\%$ then the 6 $\times$ 6 reciprocal space unit cell has a magnitude $\alpha = 13.3\%$ BZ and the $6\sqrt{3} \times 6\sqrt{3}$ reciprocal space unit cell has a magnitude $\beta = 7.7\%$ BZ. 
Peaks $v_1$, $v_2$, $v_3$ in the diffraction pattern of Figure \ref{fig:fig1}d can be written in terms of vectors along the 6 $\times$ 6 reciprocal lattice directions which are multiples of $\alpha$, added to fundamental spots. 
Point $v_1$ is separated by a vector of magnitude $\alpha$ from (0, 0), point $v_2$ is separated by a vector of magnitude 6$\alpha$ from fundamental spot G$_1$ - G$_2$ and point $v_3$ is separated by 6$\alpha$ from fundamental spot G$_2$. 
On the other hand the point $v_4$ belongs to the reciprocal lattice of the $6\sqrt{3} \times 6\sqrt{3}$ coincidence lattice at positions 5$\beta$  measured from (0, 0). 
The origin of the $v_4$ vector has been debated in the literature over long time (both in the more recent case of graphene and the older literature discussing thermal annealing of SiC to form graphite in terms of being incommensurate spots or as spots originating only from multiple scattering). Our recent high resolution SPA-LEED measurements indicate that the $v_4$ leads to strongest diffraction spots of the coincidence 6$\sqrt{3}$ lattice. 
These spots are present only when the buffer layer and first layer graphene form. 
On the other hand the spots corresponding to vectors $v_1$, $v_2$, $v_3$ are still present even when multilayer graphene is grown.

The replicas of the main Dirac cones observed  in ARPES data, described by a set of above mentioned vectors,  are shown in  \ref{fig:fig1}e.
The three replica cones originate from the three vectors $v_1$, $v_2$, $v_3$ measured from the corner of the BZ $K_2$. 
They are related to the corresponding wavevectors of the LEED pattern in Fig. \ref{fig:fig1}d if the LEED vectors are translated by the vector $\Gamma K_4$. 
The three vectors are separated by $\alpha$ (the $v_1$ vector), 1.666$\alpha$ ($v_2$) and 2.666$\alpha$ ($v_3$), respectively,  measured from $K_2$. 
The vectors $v_2$ and $v_3$ are symmetrically located about the midpoint of the side $K_2\Gamma$ and their separation is $\alpha$. 
When they are compared to the experimental ratios seen in Fig. \ref{fig:fig2} (a) and in Table \ref{table1} (normalized to the length of $K_2\Gamma$, 4.33$\alpha$) they result in ratios 0.23 for $v_1$, 0.384 for $v_2$ and 0.615 for $v_3$, which are in excellent agreement with the measured values 0.24 for $v_1$, 0.390 for $v_2$ and 0.619 for $v_3$. 
Furthermore if the vectors $v_2$ and $v_3$ are measured from the opposite corner $K_5$ of $K_2$ (by adding 4.33$\alpha$), they correspond to vectors 6$\alpha$ (4.33$\alpha$ + 1.66$\alpha$) and 7$\alpha$ (4.33$\alpha$ + 2.66$\alpha$). 
All wavevectors for the three replicas seen in the current experiments are the same as the replicas seen in ref. \cite{Nevius2015} (the first replica closest to the BZ was measured from the original corner $K_3$ so it corresponds to a separation of $\alpha$). 
In ref. \cite{Nevius2015} only the buffer layer was grown but the current work shows that they are the relevant vectors, even for much thicker graphene.
A fainter spot is seen at D4 with wavevector $v_4$ which becomes extinct in the LEED pattern after the monolayer is complete. 
The green first Brillouin zone border is the same as in Fig. \ref{fig:fig1}d.
\begin{figure}
	\centering
	\includegraphics[width=\columnwidth]{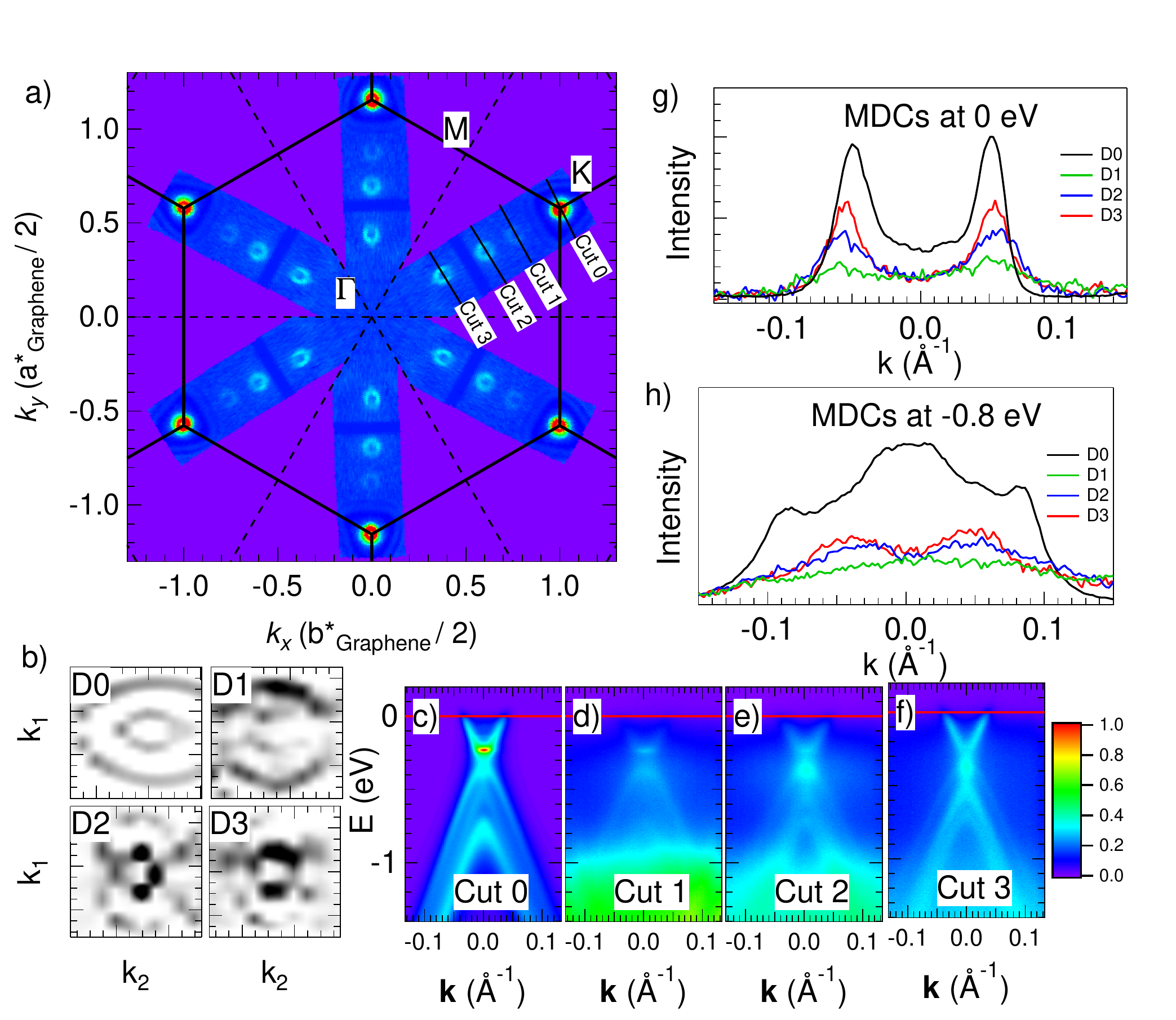}
	\caption{
	(a) Fermi surface of three-layer graphene grown on 6H-SiC substrate. 
	(b) Constant energy contour after MDC second order differentiation with respect to k at E = 0.8 eV below Fermi level of Dirac cones D0, D1, D2 and D3. $k_1$ is perpendicular to $\Gamma$ - K direction and k2 is along $\Gamma$ - K. 
	(c - f) Energy dispersion at Dirac cones D0, D1, D2. D3 in the direction perpendicular to $\Gamma$ - K.
	(g) MDCs at Fermi level in (c - f). 
	(h) MDCs at 0.8 eV below Fermi level in (c - f).}
	\label{fig:fig3}
\end{figure}

The plot of ARPES intensity at E$_F$ for a single layer graphene grown on SiC is shown in Figure \ref{fig:fig2}a and is based on a measurement over one sixth of the Brillouin zone and symmetrization. 
In addition to the ``main" Dirac cones at the corners of the BZ, there are several replicas of the main Dirac cones shifted by a set of vectors from the K point. 
Namely, there are three replicas along each symmetry line connecting the center and corners of the BZ and are located at 0.240, 0.390, 0.619 $|a_{\Gamma-K}|$ from K point. 
The $|a_{\Gamma-K}|$ = $\frac{4\pi}{3\sqrt{3}a_G}$ = 0.983 \AA\textsuperscript{-1} is the distance between $\Gamma$ point and K point in the graphene's first Brillouin zone. 
The location of each of the replicas can be constructed as a combination of the first order diffraction vectors of the graphene and SiC lattices as explained in \ref{fig:fig1}e. 
Therefore, each main Dirac cone D0 is surrounded by three sets of 6 replicas Dirac cones. 
Taking Dirac cone D0 at K$_1$ as an example, it has one replica at D1s (D1 and its 6-fold symmetry points) (vector v$_1$), two replicas at D2s (vector v$_2$ and v$_2$'), and three replicas at D3s (vector v$_3$, v$_3$' and S$_1$) which can be written as v3+v3' within first BZ. 
The obvious mechanism for creating such signal in ARPES is photoelectron diffraction, however our data demonstrates that this is not the case here: this effect is due to the weak modulation of the ionic potential in the graphene layer caused by the interface with the substrate. 
This weak modulation of the potential is ``felt" by conduction electrons as perturbation of the graphene periodic ionic potential. 
Such effect is quite important as it likely impacts the transport properties of graphene  grown on SiC substrates. 

The band dispersion along the $\Gamma$ - K symmetry direction is shown in Figs. \ref{fig:fig2}b. 
Figure \ref{fig:fig2}c-f show the band dispersion along cuts perpendicular to the symmetry axis - marked in panel (a). 
The main Dirac cone consists of a single band, a clear signature of a single-layer graphene \cite{PhysRevLett.98.206802}. 
The intensity of  each Dirac replica is significantly weaker but still clearly visible on top of the background with D3 being the strongest and D1 the weakest. 
The shapes of replica dispersions D1 - D3 are identical to D0. 
The momentum distribution curves (MDC's) at the $E_F$ and -0.8 eV are shown in panels (g) and (h), respectively. 
The separation of the MDC peaks at E$_F$ is very similar demonstrating close relation between main cone D0 and replicas D1,D2 and D3.
\begin{figure}
	\centering
	\includegraphics[width=\columnwidth]{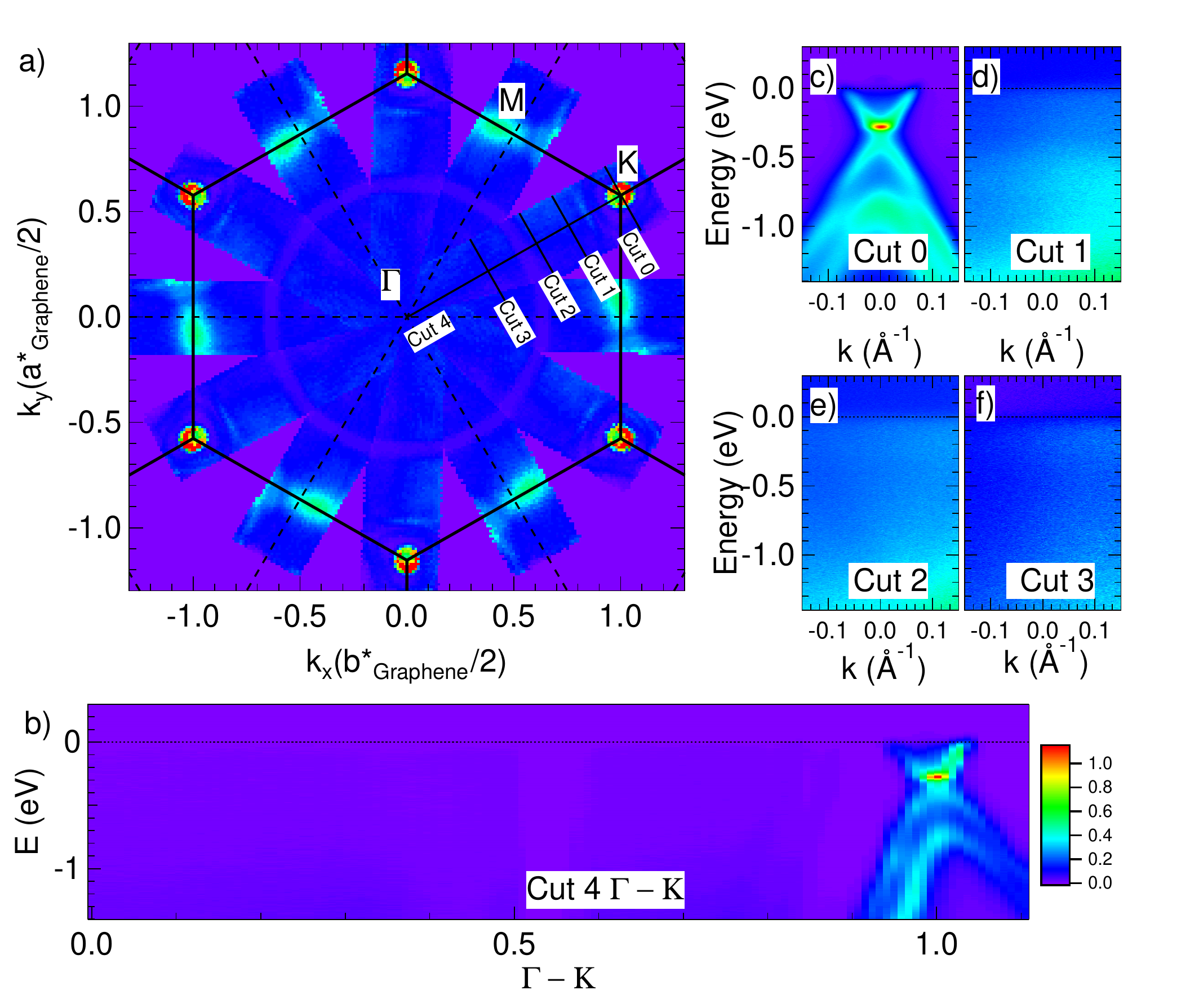}
	\caption{
	(a) Fermi surface of Dy-doped graphene on 6H-SiC substrate. 
	(b) Energy dispersion along $\Gamma$ - K direction. 
	(c - f) Energy dispersion along the directions marked in (a).}
	\label{fig:fig4}
\end{figure}

Figure \ref{fig:fig3}a shows the Fermi surface of a three-layer graphene. 
The replica cones are at the same positions as in single-layer graphene. 
Their crossections have slightly different shapes than the ones in Fig.~2(a), which may be due to additional sheets of FS originating from three atomic layers of graphene. Fig.~\ref{fig:fig3}b shows constant energy contours at 0.8 eV below the Fermi level for the main Dirac cone and the three replicas. 
The main Dirac cone and D1 have an oval crossection with long axis being horizontal (i.e. along $\Gamma$ - K ). 
This is because the vector connecting them is parallel to the $\Gamma$ - K symmetry direction. D2 and D3 cones also have an oval shape, but are oriented  perpendicular to  the $\Gamma$ - K  . 
This is because they are connected with translation vectors to the adjacent main Dirac cone that is perpendicular to the $\Gamma$ - K direction (v2,v3 in Fig.~1(e)) confirming our model shown in Fig. \ref{fig:fig1}e. 
The band dispersion for each cut along the perpendicular direction to $\Gamma$ - K is shown in \ref{fig:fig3}c - \ref{fig:fig3}f. 
There are three bands clearly visible below the Dirac point consistent with three-layer graphene \cite{PhysRevLett.98.206802}. 
The overall intensity of the Dirac cone replicas are similar to the ones in single-layer graphene, signifying that they are not due to photoelectron diffraction. 
If that would be the case, one would expect the replicas to be much weaker in tri-layer graphene. 
Definitive evidence for intrinsic origin of the Dirac cones can be directly seen in the relative intensities of the three bands. 
In each of the dispersion data shown in Figs.~\ref{fig:fig3}c - \ref{fig:fig3}f, the pattern of the intensities of each band is different. 
In D0, the inner and outermost bands are more intense below the Dirac point and there is also very strong intensity above that point. 
D1 has very weak intensity above the Dirac point and weaker inner band below. 
D2 is similar to D1, but here the two inner bands below the Dirac points are strongest. 
On the other hand, D3 has strong intensity above the Dirac point, whereas below the middle band is the most visible. 
If the replicas of the Dirac cone had originated from photoelectron diffraction, the pattern of the intensities would exactly match the one seen in the main Dirac cone. This is because the scattering form factors are expected to vary slowly with scattering vector and should be independent of small variations of photoelectron kinetic energy. This is clearly not the case here. 
Since each of the four Dirac cones D0-D3 shown in Figs.~3(c-f) corresponds to a different wavevector (as defined earlier  by multiple of $\alpha$=13.33\% of BZ), the observed intensity differences confirm that the Dirac cone replicas we report here are intrinsic to graphene. 
They are due to  weak modulation of the ionic potential experienced by the graphene electrons caused by the Moir\'e pattern that forms at the interface of the SiC substrate and graphene. 
Further evidence that the Dirac cone replicas are related to the interface Moir\'e can be seen in Fig.~4. Monolayer graphene was intercalated after deposition of ~1ML of Dy  at room temperature followed by heating   to 800$^{\circ}C$. 
A primary bilayer Dirac cone is still present as seen in Fig.~4(a) judging from the number of visible bands, except that there is additional intensity at the M point. 
More remarkably, no replica cones are seen away from the K-point - Figs. 4(d-f). 
It  is known that the intercalated Dy is bonded between graphene and the SiC substrate, converting the buffer to graphene and therefore the monolayer to bilayer graphene. 
The intercalation destroys the Moire modulation and the component of the ionic potential responsible for the replica cones at the LEED wavevectors of Fig.~1(d).  
This supports the  previous conclusion for the pristine sample  that the replica cones are  not due to photoelectron diffraction.
 
\section{conclusion}

In summary, we report the presence of additional features in the electronic structure of graphene grown on SiC substrates. 
Namely there are three sets of replica cones of the main Dirac cones at wavevectors observed in the LEED patterns that are expressed as  linear combinations of the reciprocal vectors of graphene and the SiC substrate. 
We also demonstrate that these replica cones are intrinsic rather than due to the photoelectron diffraction process because they exist in single and tri-layer graphene and the pattern of intensities ( of the bands measured in the ARPES spectra) is very distinct from the ones present in the main Dirac cone. 
This indicates that the ionic potential experienced by the graphene electrons is modulated with these additional periodicities which originate at the graphene-SiC interface. 
From these combined LEED, STM  and ARPES experiments we demonstrate that the interface of 2-d material can only be understood when real space structure and band structure are measured in parallel. The current experiments show that the wavevectors of the three replica cones   observed in ARPES spectra with different  intensity modulation are correlated to the real space Moire periodicity deduced from the LEED patterns and STM images. The presence of these features and the modulation should play a key role in understanding recent transport measurements. 

This work was supported by the US Department of Energy, Office of Basic Energy Sciences, Division of Materials Sciences and Engineering. Ames Laboratory is operated for the US Department of Energy by the Iowa State University under Contract No. DE-AC02-07CH11358. 

\bibliography{Graphene}

\end{document}